\newcommand{\beq}{\begin{equation}}
\newcommand{\eeq}{\end{equation}}
\newcommand{\Xmax}{X^{\mu}_{\rm max}}
\newcommand{\gcm}{{\rm g\,cm^{-2}}}
\newcommand{\Conceicao}{{Concei\c{c}\~{a}o}}

\newcommand{\QII}{QGSJET-II.03}

\documentclass[twocolumn]{article}
\usepackage{graphicx} 

\title{Extensive Air Showers: from the muonic smoking guns to the hadronic backbone}

\begin{document}

\author{L.~Cazon$^1$ \\
$^1$ LIP, Av. Elias Garcia 14-1, 1000 Lisboa}

\twocolumn[
  \begin{@twocolumnfalse}
    \maketitle
    \begin{abstract}
 Extensive Air Showers are complex macroscopic objects
 initiated by single ultra-high energy particles. They are the result of millions of high energy reactions in the atmosphere and can be described as the superposition of hadronic and electromagnetic cascades. The hadronic cascade is the air shower backbone, and it is mainly made of pions. Decays of neutral pions initiate electromagnetic cascades, while the decays of charged pions produce muons which leave the hadronic core and travel many kilometers almost unaffected. Muons are smoking guns of the hadronic cascade: the energy, transverse momentum, spatial distribution and depth of production are key to reconstruct the history of the air shower. In this work, we overview the phenomenology of muons on the air shower and its relation to the hadronic cascade.  We briefly review the experimental efforts to analyze muons within air showers and discuss possible paths  to use this information.
\vskip1.5cm
    \end{abstract}
\end{@twocolumnfalse}
]
\section{Introduction}
\label{intro}

Our understanding of high energy physics is supported by experiments up to the TeV scale. Beyond such high energy frontier we must rely on extrapolations of our theories in terrains which might hide unexpected phenomena. The only direct processes surpassing these high energies and which might represent a challenge in particle physics are the reactions initiated by Ultra High Energy Cosmic Rays (UHECR) high up in the atmosphere. They reach up to $\sim 10^{20}$ eV in the lab system, which corresponds to a center of mass energy of about $\sim 400$ TeV.

The origin and nature of UHECR remains a mystery. Our current understanding says that the vast majority of these particles are hadronic (atomic nuclei \cite{Abraham:2010yv,Cazon:2012rk}), excluding neutrinos \cite{Abraham:2009uy} and photons \cite{Abraham:2009qb}. The final solution to the UHECR puzzle must put together different pieces: the astrophysical mechanisms that allow the acceleration to such gigantic energies, the propagation through the intergalactic space filled with magnetic fields, and last, the subject of this paper, the interaction with the Earth's atmosphere, which creates Extensive Air Showers (EAS).

EAS encode the information of the primary among millions of secondaries by means of high energy interactions which lie on kinetic regions never accessed by experiments before. Among all secondaries, muons can travel many kilometers from the hadronic backbone almost unaffected, carrying valuable information.  Understanding this information is key to break the degeneracy between the uncertainties on the extrapolation of the hadronic interaction models to the highest energies and the composition of the UHECR {\it beam}.

This paper is organized as follows: In section \ref{sec-2} we overview certain aspects of air showers and discuss the energy balance between the hadronic and electromagnetic cascade. In section \ref{sec-3} we discuss how muons are produced in the hadronic cascade inheriting valuable information from it. In section \ref{sec-4} we illustrate the ground distributions of muons. In section \ref{sec-5} we briefly discuss the experimental efforts to use this information. In section \ref{sec-6} we conclude.

\section{Extensive Air Shower dynamics}
\label{sec-2}

Extensive Air Showers are complex phenomenon initiated by a single particle with an enormous energy. The collision with an air nucleus generates typically thousand of secondaries, which can interact again, creating a multiplicative process which is referred as cascade, and that can reach up to $10^{11}$ particles at ground level for $10^{20}$ eV showers.

Depending on the kind of particles driving the multiplicative process, there are two main subtypes of cascades. The ones initiated and driven by photons or electrons, and the ones originated and driven by hadrons.

The study of the cascade can be done by means of the cascade equations, assuming some simplifications, or by means of full Monte Carlo simulations that include many important details difficult to account for otherwise.
On the other hand, Heitler models offer a simplified version of the main multiplicative process of a cascade and serves to qualitatively understand the most important features, giving approximated values for relevant variables of the cascade. See for instance \cite{Matthews:2005sd} for more details on the hadronic and EM cascade. 

\subsection{The electromagnetic  and the hadronic cascades}
When a high energy photon is injected into matter, the most likely process to occur is an electron-positron pair production. Each of the new particles suffers bremsstrahlung, producing new photons. This multiplicative process repeats itself $n$ times originating the so called electromagnetic (EM) cascade. The total number of particles grows  as $2^{n}$. The energy of secondaries decreases as $E=\frac{E_0}{2^n}$ to eventually reach the so called critical energy ($E_c$ $\sim$ 80 MeV) at which electrons are more likely to lose their energy through ionization. At this point the cascade reaches the maximum. After that, the multiplicative process stops, and the number of particles declines.
 The EM cascade practically keeps all the energy flowing within the EM channel, and does not leak into the hadronic cascade except for a small fraction by photopion production.

\begin{figure}[!ht]
  \begin{center}
    \includegraphics[width=7cm]{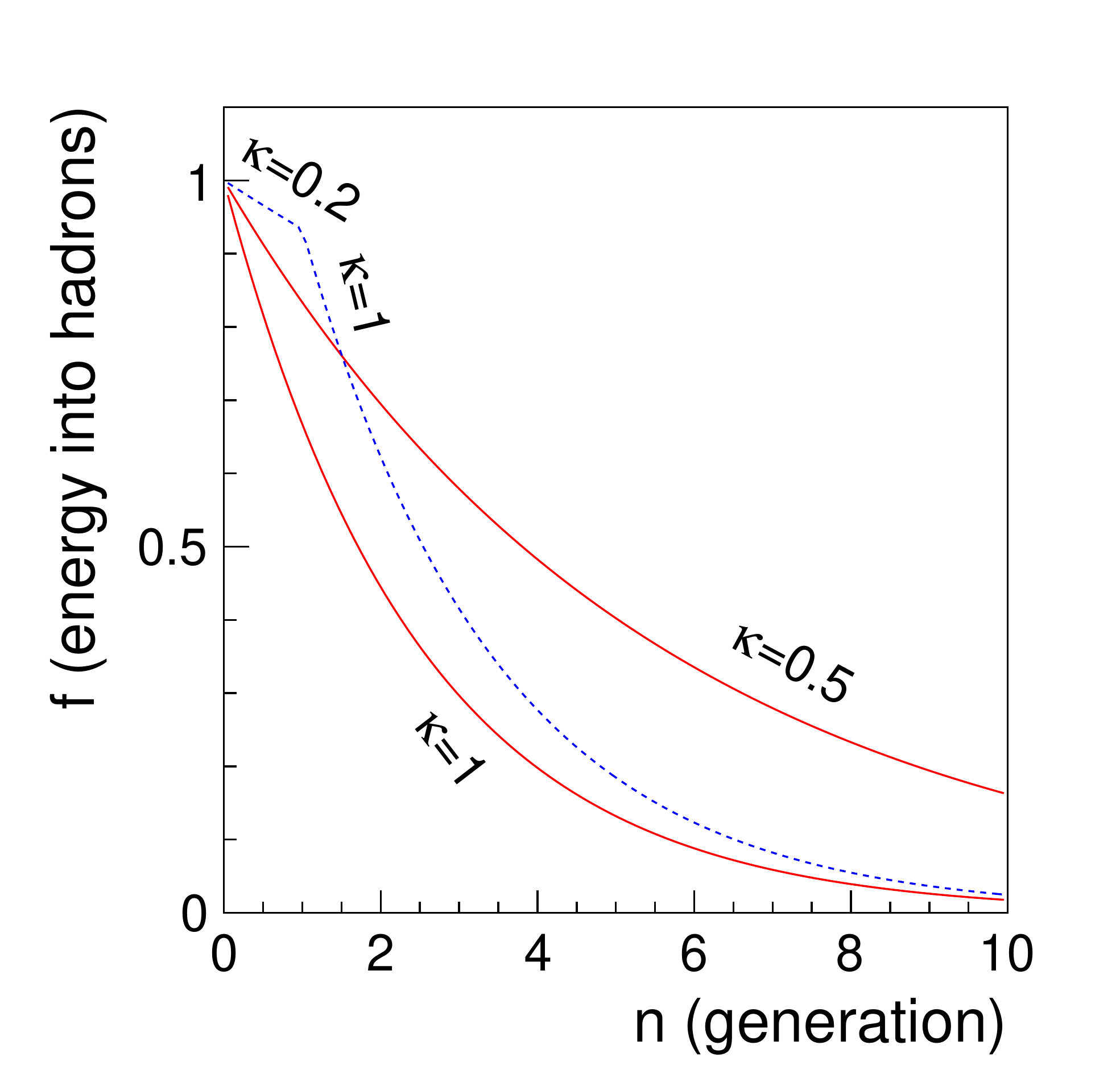}
    \caption[]{Energy fraction evolution with generation $n$.}
    \label{f:fE}
  \end{center}
\end{figure}

On a hadronic reaction at high energies,  $\sim$ 80\% of the produced particles are pions, ($\pi^+$, $\pi^-$ and $\pi^0$) in a $\sim$ 1:1:1 ratio, and $\sim$ 8\% are kaons, ($K^0_L$, $K^0_S$, $K^+$ and $K^-$) also with a $\sim$ 1:1:1:1 ratio. Neutrons/protons are produced with an overall probability of $\sim$ 4-5\%, and the rest is shared among other particles at the subpercent level as given by QGSJET-II.03  \cite{AlvarezMuniz:2012dd}\cite{Ostapchenko:2005nj}.

Neutral pions feed the EM cascade almost immediately, whereas charged pions either interact, sustaining the hadronic cascade, or decay into muons (99.988\%). In the same way, kaons interact feeding the hadronic cascade untill they reach their critical energy, which is of the same order of magnitude compared to pions. $K^0_S$ has a shorter lifetime ($c\tau=4$ cm) compared to the rest of kaons ($c\tau\sim$ few meters), which implies a higher probability of decay before interacting. A few hadronic generations after the first interaction, $K^0_S$  decays, 31\% of the times  into $\pi^0 \pi^0$, and 69\% into $\pi^+\pi^-$. This means that kaons go from a $\sim$ 0\% contribution to the EM cascade in the first generations up to $\sim$ 8\% in higher generations, compared to the steady $\sim$ 33\% contribution of pions to the EM cascade. Finally, neutrons and protons keep interacting hadronically with no direct feeding into the EM cascade.

The most relevant features of  hadronic shower can also be approximately described by a Heitler model.
After each hadronic generation $n$, there are created  $m$ particles which subdivide in two main categories: those which continue to feed the hadronic cascade, and those which feed the EM cascade, leaving the hadronic channel. They typically correspond to charged and neutral pions, in a 2/3$m$ and 1/3$m$ proportion. Thus, total number grows with the hadronic generation as $(2/3m)^{n}$  whereas the energy decreases as $E^{\pi}=\frac{E_0}{m^n}$. The energy fraction $f$ carried by the sum of all charged pions in generation $n$ to the total shower energy $E_0$ is
\begin{equation}
f=\frac{\sum E^{\pi}}{E_0}=\left(1-\frac{1}{3}\right)^{n}
\end{equation}
That is, in each generation, the energy carried by charged pions $\sum E^{\pi}$ is reduced by a factor $\frac{2}{3}$. In a more realistic approach, we can include an effective factor $\kappa\in[0,1]$ that modifies the amount of energy flowing to the EM cascade through $\pi^0$ decay as:
\begin{equation}
f= \left(1-\frac{1}{3}\kappa \right)^{n}
\end{equation}
$\kappa$ can account for different aspects of the hadronic reactions. For instance, if a leading baryon takes $(1-\kappa)E_0$, $\kappa$ accounts for the inelasticity, being the fraction of energy going into pion production, and therefore $\frac{1}{3}\kappa E_0$ goes into the EM channel, as explained in \cite{Matthews:2005sd}. There might be other mechanisms that could effectively reduce the feeding to the EM channel, for instance, increasing the amount of kaon production \cite{AlvarezMuniz:2012dd}.

\begin{figure*}[!Ht]
\begin{center}$
\begin{array}{cc}
\includegraphics[width=7cm]{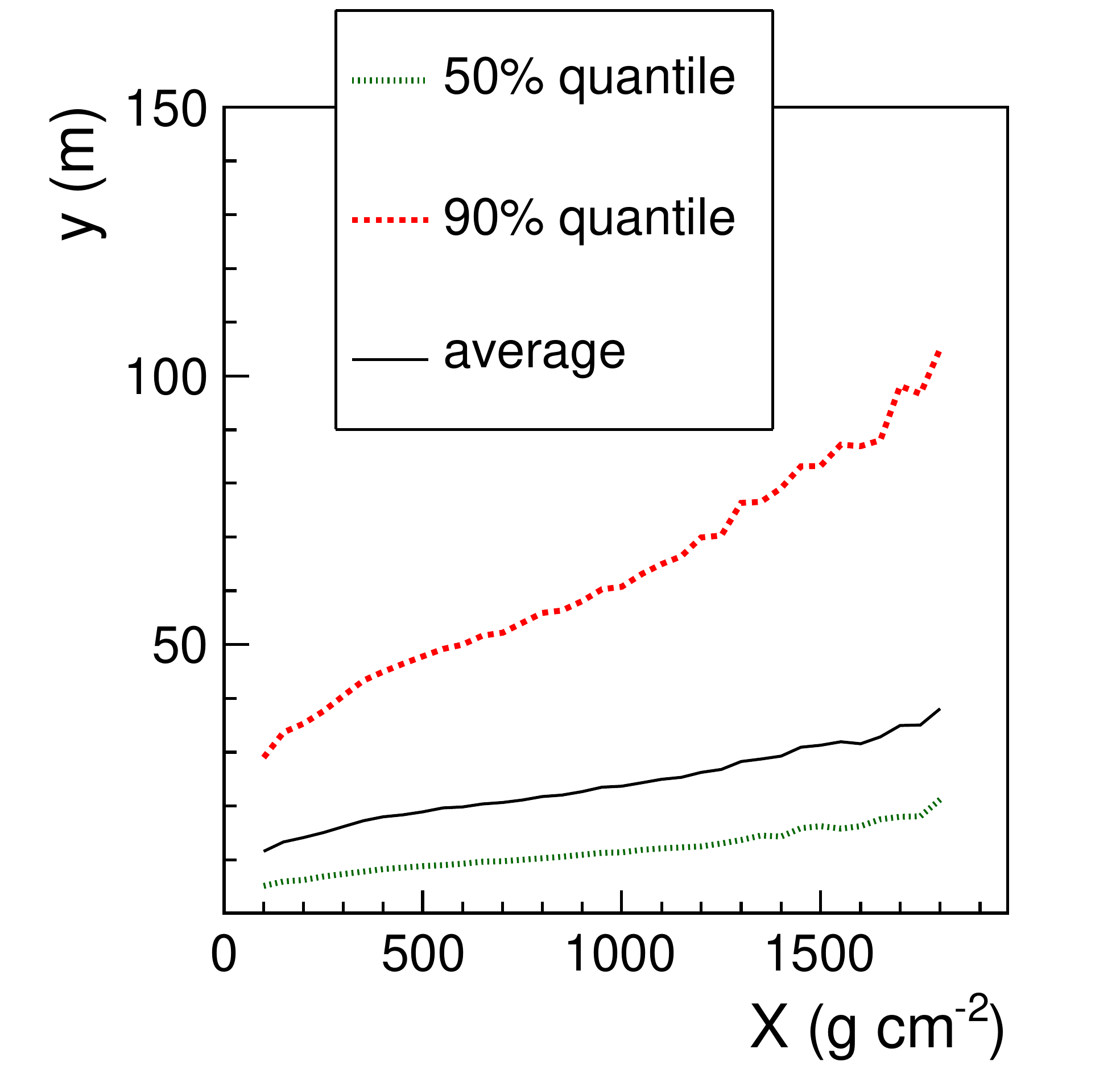}&
\includegraphics[width=7cm]{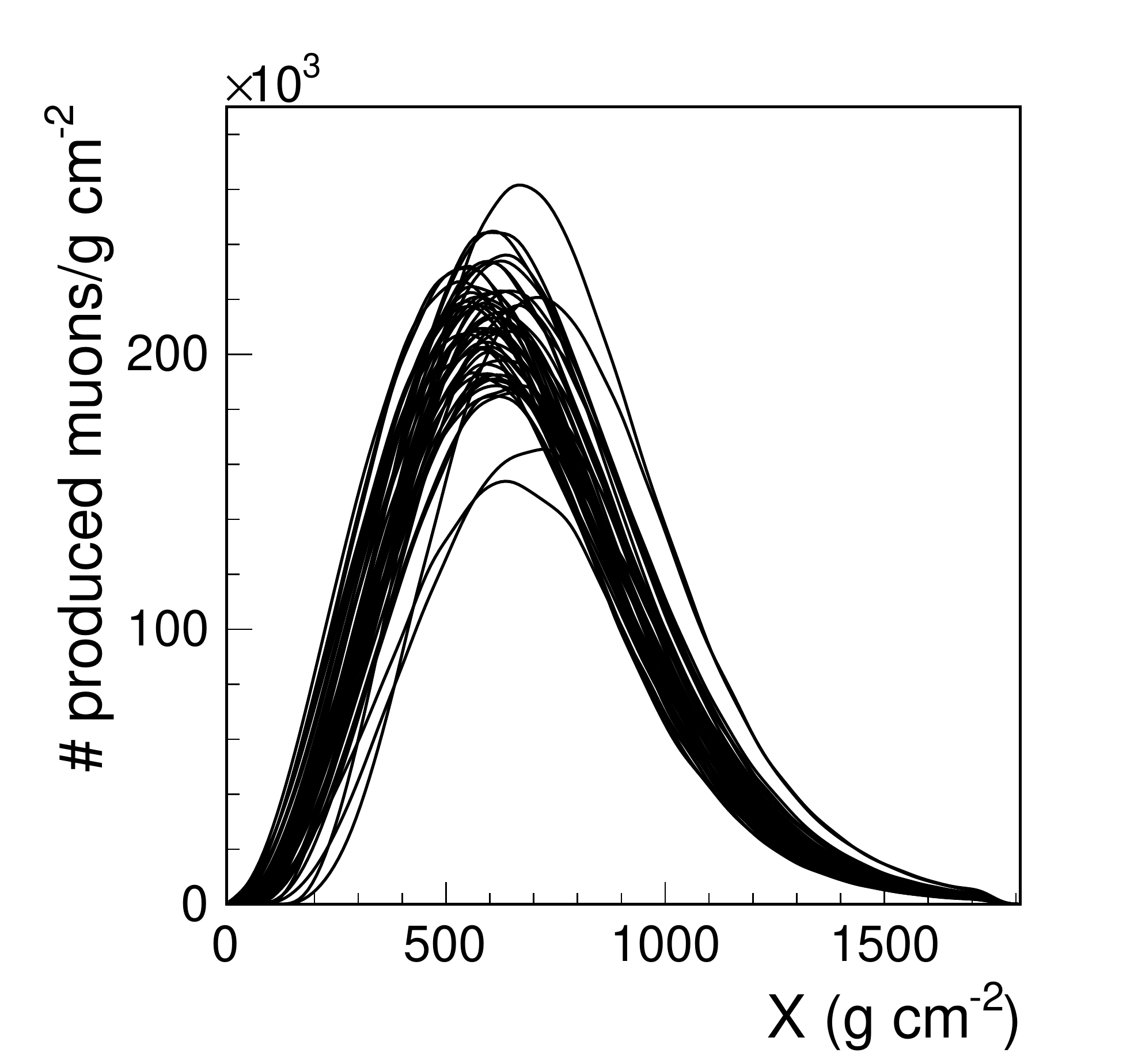}
\end{array}$
\end{center}
\caption{Left panel: average, median and 90\% quantiles of the $y$-distribution for different depths. Right panel: total number of muons produced per $\gcm$, $h(X)$, for 50 proton showers at $10^{19}$ eV and 60 deg. 
} 
\label{f:hX}
\end{figure*}

\subsection{The energy balance between cascades}

The energy share between both cascades evolves with the hadronic generation as showed in Fig. \ref{f:fE}. In the beginning all the energy is in the hadronic sector. After 3 generations, ($\kappa=1$) 70\% of the energy has been transfered to the EM sector. This means that the evolution of the EM cascade is rapidly decoupled from the hadronic cascade. Also shown is the case for $\kappa=0.5$, where the transfer from the hadronic to the electromagnetic cascade is slower. The energy balance affects the longitudinal developement and the muon content of the shower, see for instance \cite{AlvarezMuniz:2012dd}. 

The factor $\kappa$ can change with the energy of the hadronic reaction, and thus change with the  generation $n$. The energy at which the first and second generation reactions occur might be out of reach of the current man made accelerators. Fig. \ref{f:fE} also shows a case where $\kappa$ changes from a value $\kappa=0.2$ to  $\kappa=1$ after the first generation. It can be seen how the energy balance of the whole shower is affected.

\section{The production of muons in EAS}
\label{sec-3}

Most muons in the shower come from the decay of pions, which are 10 times more numerous than kaons. Kaon decay can lead directly to muons (20\%) or to charged pions (40\%).

Simple kinematics shows that the maximum transverse momentum $p_t$ that muons can obtain 
is just the center of mass momentum of the outgoing particles, which is 29.8 MeV. Given that the total momentum of the parent particles is of the order of a few tens of GeV, the direction of motion hardly varies, with 
deviation angle $\Delta \theta_{\pi \mu} \sim 0.01^\circ$.

The experimental data of hadronic collisions available up to a few hundreds of GeV per nucleon in the center of mass
show a $p_t$ distribution that decreases exponentially $\frac{dN}{2 \pi p_t dp_t} \propto \exp(-\frac{pt}{Q})$
where $Q$ changes slowly with the energy of the collision and the rapidity
region. $Q$ is of the order of tenths of GeV/c, that compared to the muon
maximum transverse momentum available from the pion decay ($\sim$ 0.03 GeV)
gives a 10\% correction. This makes the $p_t$ distribution of the outgoing muons
 very similar to that of their parents. This is a very important feature
responsible for many of the observed characteristics of the hadronic and
muonic showers.

In \cite{Cazon:2003ar}, it was argued that the transverse position of the production of muons, thus of the parent mesons decay, is confined to a relatively narrow cylinder: as the angle with respect to the shower axis goes as $\sin \alpha = \frac{cp_t}{E}$, the average traveled distance before the pion decay is $l=\frac{E}{m_{\pi}c^2} c \tau_\pi$, where $\tau_\pi$ and $m_\pi$ are the lifetime and mass of the charged pions. The perpendicular distance to the shower axis of the pion decay is $r_{\pi}=l \sin \alpha =\tau_\pi 2 Q /m_\pi \sim 22$ m.

Note that after each interaction $n$, the $p_t$ increases as $p_t \sim Q\sqrt{n}$. The outgoing angle goes as a geometrical progression with $n$ as
$\alpha_i \simeq \sin \alpha_n=\frac{Q \sqrt{N}}{\left(3/2 N_{ch} \right)^n}$. The total outgoing angle $\sum^n_{i=1} \alpha_i$ is then dominated by the last interaction 
\begin{equation}
\sum^n_{i=1} \alpha_i \simeq \frac{Q \sqrt{N}}{\left(3/2 N_{ch} \right)^n}
\end{equation}

 Fig. \ref{f:hX} (left panel), displays the $y$-coordinate\footnote{We use a system of coordinates (cylindrical $(r,\zeta,z)$ or Cartesian $(x,y,z)$, as convenient) with the $z$-axis aligned with the shower axis and (0,0,0) being at ground level.} (the shower axis is at $y$=0) containing 50\% and 90\% of the production points as a function of the atmospheric depth. Also displayed is the average value, which is of tens of meters. This distance is small when compared to the distances involved in EAS experiments, which span from hundreds of meters to several kilometers in the perpendicular plane. For instance, the Pierre Auger Observatory has its tanks separated by 1.5 km \cite{Abraham:2004dt}. Therefore, the position where the muon has been produced can be approximated by $(0,0,z)$, or simply $z$.

Every $dX$\footnote{$X$ is used as being equivalent to $z$ corresponds to the column-air along $z$, $X=\int_z^\infty \rho(z') dz'$}  along the shower axis, $dN$ muons are produced within a given energy and transverse momentum interval $dE_i$ and $dp_t$. Their overall distribution at production can be described in general with a 3-dimensional function, as:
\beq
\frac{d^{3} N}{dX\, dE_i\,dcp_{t}}= F(X,E_i,cp_{t})
\label{eq:totaldistribution}
\eeq

The projection into the $X$ (or $z$) axis becomes
\beq
h(X)=\int F(X,E_i,cp_{t}) dE_i dcp_t
\eeq
and it is the so called {\it total/true} Muon Production Depth (Distance) distribution, or MPD-distribution for short.  It does not depend on the observational conditions since it does not contain any propagation effects of muons through the atmosphere. A detailed study of its shape is done in \cite{Andringa:2011ik}.
 Notice that this is different from the MPD-distributions of detected muons at a given position on ground  $\frac{dN}{dX}|_{(r,\zeta)}$, which includes the effects of propagation, as it will be explained later. This distribution is sometimes referred to as {\it apparent} MPD-distribution.
\begin{figure*}[!Ht]
  \begin{center}$
\begin{array}{cc}
    \includegraphics[width=7cm]{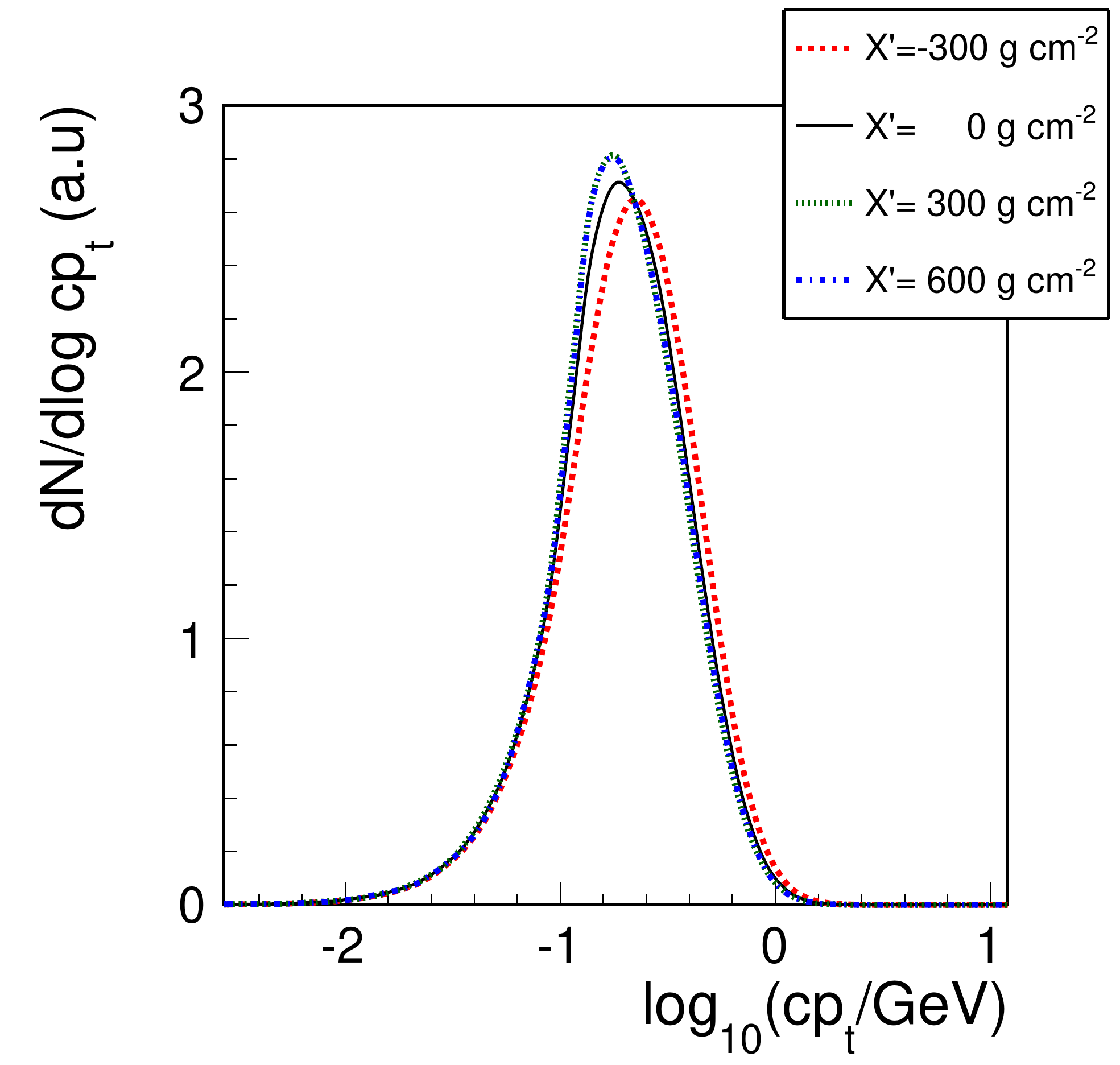}&
    \includegraphics[width=7cm]{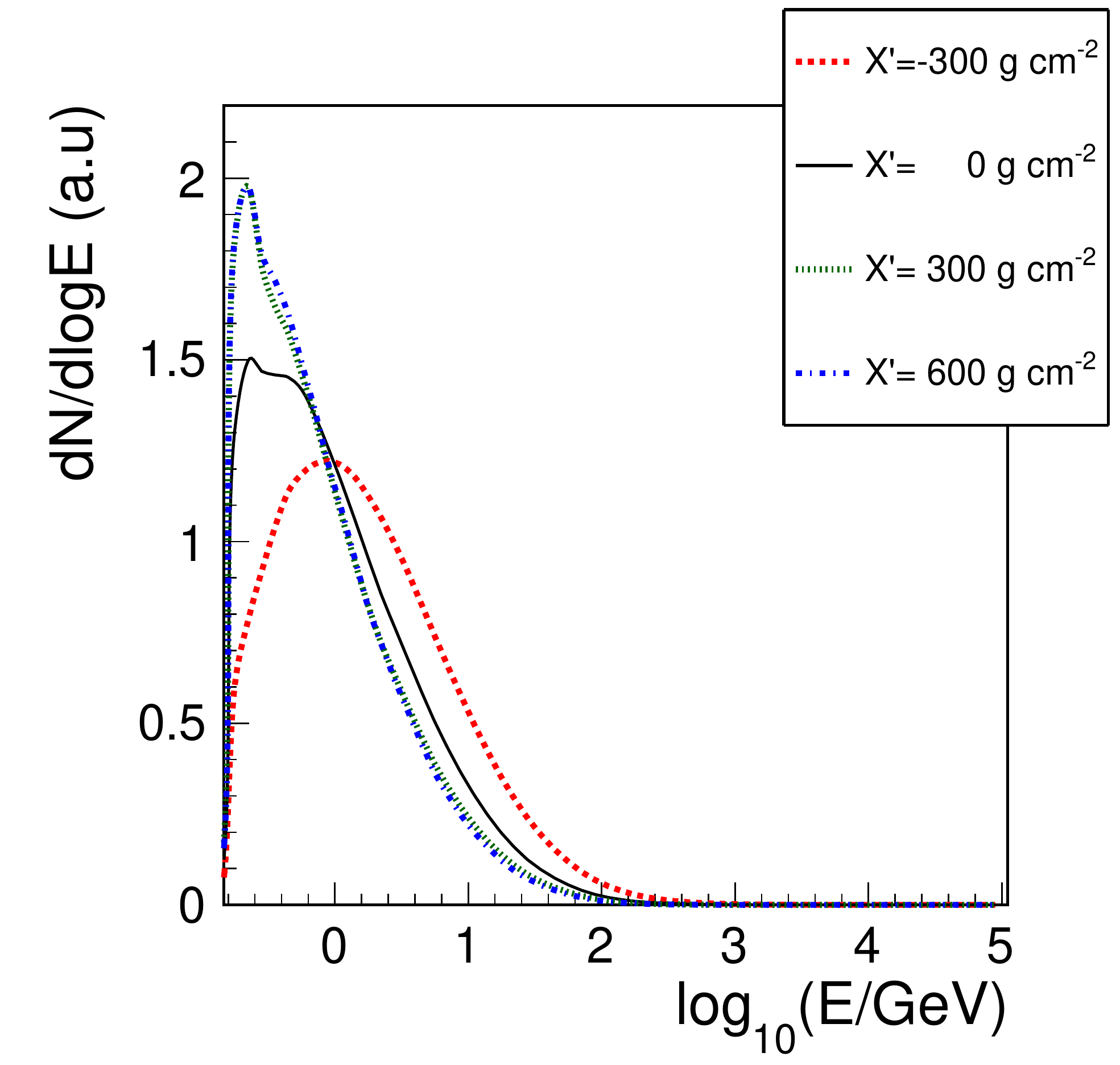}
\end{array}$
    \caption[]{Normalized average energy (left panel) and average $p_t$ (right panel) distribution of all muons at production for proton showers at $10^{19}$ eV and 60 deg zenith angle simulated with \QII  ~ at different $X'$ layers.}
    \label{f:Ept}
  \end{center}
\end{figure*}

The total number of muons produced in a shower is
\beq
{\cal N}_0=\int h(X) dX
\eeq
It should be noted that this number is intrinsically different from the number of surviving muons, which is affected by the fluctuations of the depth of the first interaction, and thus change the distance traveled by muons to the ground. Some of the techniques used by experiments like Auger \cite{Schmidt:2007vq} use a fixed distance to the shower core, so they can also be affected by the lateral spread of the parent mesons. 

Eq. \ref{eq:totaldistribution} can be factorized and expressed as the product
\beq
F(X,E_i,cp_{t})= h(X) \,f_X(E_i,cp_t)
\label{eq:fX}
\eeq
where the function $f_X(E_i,cp_t)=\frac{F(X,E_i,cp_{t})}{h(X)}$ becomes the normalized $E_i$ and $cp_t$ distribution at a given production depth $X$. In the approximations made in \cite{Cazon:2003ar,Cazon:2004zx,PhDCazon}, $f_X$ did not depend on $X$ and it was factorized in 2 independent distributions on $E_i$ and $cp_{t}$. This allowed analytical approximations of the distributions at ground. In \cite{Cazon:2012ti} we have included these correlations, improving the accuracy of the energy, production depth, and time distributions at ground, and allowing for a proper description of the muon lateral distribution at ground.

The function $h(X)$ tracks the longitudinal development of the hadronic cascade and represents the production rate of muons per $\gcm$. Its shape and features are extensively discussed in \cite{Andringa:2011ik}. The depth at which $h(X)$ reaches the maximum is denoted as $\Xmax$. 
 $\Xmax$ correlates with the first interaction point $X_1$ which corresponds to the first interaction of the primary in the atmosphere and the start of the cascading process \cite{Andringa:2011ik}. The most important source of fluctuations in air showers corresponds to the fluctuations of $X_1$, which causes an overall displacement of the whole cascade at first approximation. The amount $X'\equiv X-\Xmax$  defines the amount of traversed matter with respect to the shower maximum. The distributions can be expressed in terms of $X'$, where the most important source of fluctuations has been eliminated, and only the remaining effects are present.

In Fig. \ref{f:hX} (right panel) $h(X)$ is shown for a sample of 50 showers. The fluctuations on the normalization and on $\Xmax$ are clearly observed. 
In \cite{Cazon:2012ti} it was shown that both the energy and the transverse momentum show similar features when referred to the same distance to the shower maximum, $X'$. 

In   \cite{Cazon:2003ar,Cazon:2004zx,PhDCazon} the muon spectrum at production was approximated by a power law, $E_i^{-2.6}$, following the high energy tails of the pion production on hadronic reactions.  A more accurate description of this the spectrum was done in \cite{Cazon:2012ti}: at low energies the single power law clearly does not work and, in addition, the energy spectrum evolves with $X'$ by becoming softer, and stabilizing the shape after the shower maximum. In Fig. \ref{f:Ept}, left panel, the average energy spectrum of all muons at production is displayed for proton showers at $10^{19}$ eV in different $X'$ layers.

The transverse momentum distributions are responsible for most of the lateral displacement of muons with respect to the shower axis. In \cite{Cazon:2003ar,Cazon:2004zx,PhDCazon}, the $p_t$ distributions were approximated by an unique function,  $dN/dp_t=p_t/Q^2 \exp(-p_t/Q)$, independent of the energy of the muon and its production depth, primary mass and zenith angle. In \cite{Cazon:2012ti}, we uncover in detail all the dependencies.
 As the shower evolves, the $p_t$ spectrum becomes softer (Fig. \ref{f:Ept}, left panel shows the evolution as a function of $X'$).  Besides this dependence on $X'$, the $p_t$ distributions also depend on the energy of the muons, as discussed in \cite{Cazon:2012ti}.  The low energy muons display a smaller $p_t$, and at high energies, the $p_t$ distribution prefers higher $p_t$ values. We have found that the different correlations of the $p_t$ with $E_i$ and $X$ must be included into the model in order to properly predict the muon lateral distribution at ground.

In \cite{Cazon:2012ti} it is shown that there are mild dependencies of both the energy and $p_t$ distributions on the energy and zenith angle of the primary. In addition, the photon initiated showers display quite different distributions due to the different nature of the processes that lead to the muon production, through photopion production. Proton and iron showers, and different hadronic models also display mild differences among them.

\section{Propagation and ground distributions}
\label{sec-4}
 In \cite{Cazon:2012ti} it was shown that a few simple considerations are enough to account for most of the features observed in the muon distributions at ground.

Firstly, muons exit the shower axis with an angle $\alpha$ determined by the energy and transverse momentum of the muon at production ($\sin \alpha = \frac{cp_t}{E_i}$). The polar angle is distributed symmetrically over $2\pi$. Once the muon is produced, the trajectory is extrapolated in a straight line to the ground, and the arrival time due to geometric path is calculated. Once the main trajectory is defined, the energy loss, decay probability, multiple scattering and effects of the magnetic field are accounted for and the impact point on ground and arrival time delay are corrected. 

Table \ref{tab-1} summarizes different effects for 5 GeV and 10 GeV muons produced at $z=10$ km and arriving at a distance from the core $r=1000$ m.
 
The most important propagation effects that shape the ground distributions are, in this order: geometry, decay and energy loss. The magnetic effects become more important in showers with zenith angle above $60$ degrees. On the other hand, the multiple scattering effects are negligible at distances to the core above 100 m.

\begin{table}
\centering
\label{tab-1}    
\begin{tabular}{lll}
\hline
Energy at production & 5.0 GeV & 10.0 GeV  \\
Energy at ground& 3.0 GeV & 7.8 GeV \\\hline
Probability of survival & 0.67 & 0.84 \\\hline
Geometric delay & 165 ns & 165 ns \\\hline
Kinematic delay & 12 ns & 2.3 ns \\
Geomagnetic delay & 0.04 ns & 0.01 ns\\
MS time delay & 1.5 ns & 0.8 ns \\\hline 
Geomagnetic lateral deviation & 83 m & 17 m \\\hline
MS lateral smearing & $\sim$ 60 m & $\sim$ 35 m \\\hline 
\end{tabular}
\caption{Summary of the different effects after propagation for a muon produced at $z$=10 km and arriving at r=1000 m at 60 deg zenith angle, and geomagnetic field strengh perpendicular to the shower axis $B_\perp=$ 10 $\mu$T (MS stands for Multiple Scattering).}

\end{table}

\subsection{The energy distribution}

The energy at ground $E_f$ was analyzed as a function of the impact point on ground $(r,\zeta)$. Typically, the muon energy is not directly measured by cosmic ray detectors since it would require carpeting extensive areas with particle detectors like those used in accelerator experiments. Nevertheless, the spectrum of muons has an impact on other quantities that are measured by current air shower detector arrays, like the muon lateral distribution at ground, the arrival angle, and the arrival time delay.

\begin{figure}[!h]
  \begin{center}
   \includegraphics[width=7cm]{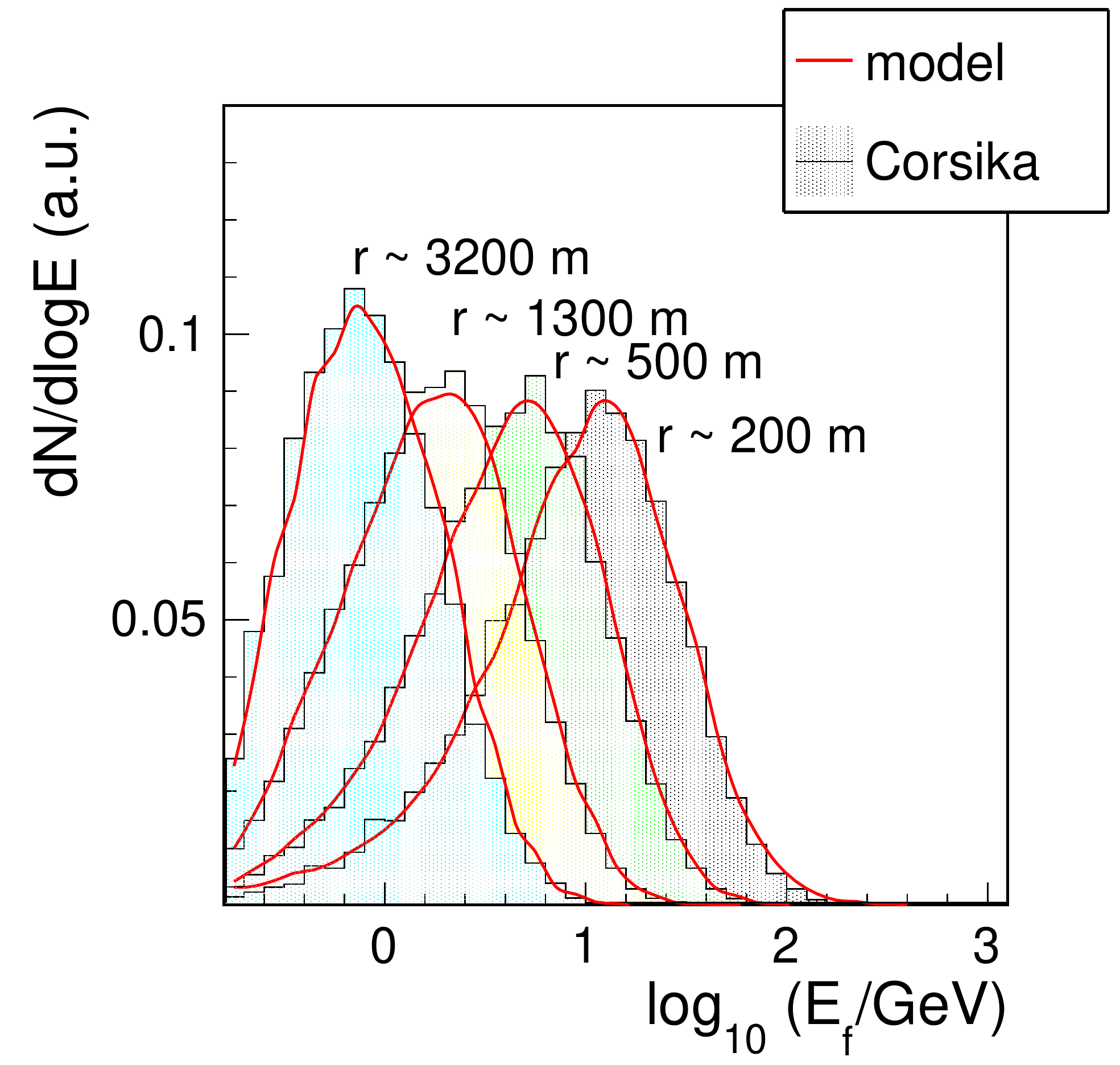}
    \caption[]{Normalized energy spectrum of muons arriving at ground for a 60 deg shower at different distances from the core as given by CORSIKA compared to the prediction of the model. }
    \label{f:Eground}
  \end{center}
\end{figure}

Fig. \ref{f:Eground} displays  the normalized energy spectra of a 60 deg shower, at different distances from the shower core. The energy of muons decreases as $\sim 1/r$ and increases with the zenith angle  \cite{Cazon:2003ar,Cazon:2004zx,PhDCazon}, being the details determined by the $p_t$, $z$ and $E_i$ distributions. Low energy muons dominate at large distances from the core.

\subsection{Apparent production depth distribution} The shape of the production depth distribution of the detected muons, the {\it apparent} MPD-distribution, changes with the observation position. The angular position of the observation point respect to the production point $z$, selects particular $(E_i,p_t)$ regions which can be more or less populated. In addition, the propagation effects, specially the decay, modulate the {\it apparent} MPD-distribution depending on the energy spectrum of muons and also the path traveled from production to ground, $l$.
\begin{figure}[!h]
  \begin{center}
    \includegraphics[width=7cm]{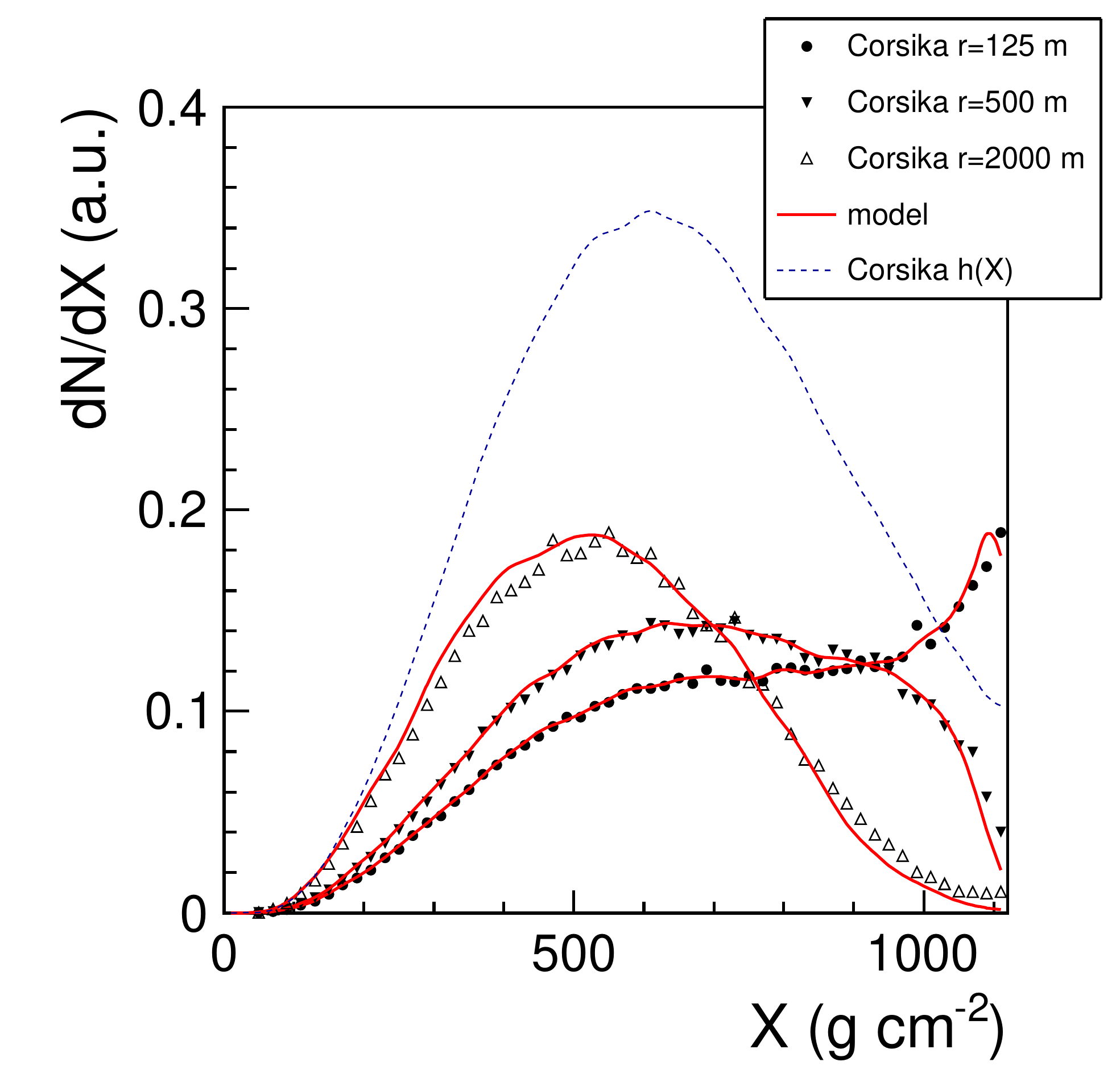}
    \caption[]{Comparison of several {\it apparent} MPD-distributions, $dN/dX|_{(r,\zeta)}$, for a 40 deg shower at different distances from the core. The {\it total/true} MPD-distribution ($h(X)$) is also plotted for comparison. Normalizations are arbitrary.}
    \label{f:dNdX}
  \end{center}
\end{figure}
Fig. \ref{f:dNdX} displays the {\it apparent} MPD-distributions for a 40 deg shower at different distances from the core, where the distortions introduced in the $dN/dX|_{(r,\zeta)}$ distributions when compared to $h(X)$ can be clearly observed.
 
The $dN/dX|_{(r,\zeta)}$ distribution is never directly observed, but reconstructed from the arrival time or the arrival angle at ground. The correct inference of the {\it total/true} MPD-distribution, $h(X)$, requires the knowledge of the exact dependence of $dN/dX|_{(r,\zeta)}$ with the observation point coordinates and detection energy threshold. $dN/dX|_{(r,\zeta)}$ explores different kinematic regions at production when reconstructed at different distances from the core. For instance, the algorithm proposed in  \cite{Cazon:2004zx} and \cite{GarciaGamez:2011} requires the conversion of each $dN/dX|_{(r,\zeta)}$ observed in each station to an universal distribution in order to sum up the contributions of all detectors in a single shower.

\subsection{Time distributions}

\begin{figure*}[!Ht]
\begin{center}$
\begin{array}{cc}
\includegraphics[width=7cm]{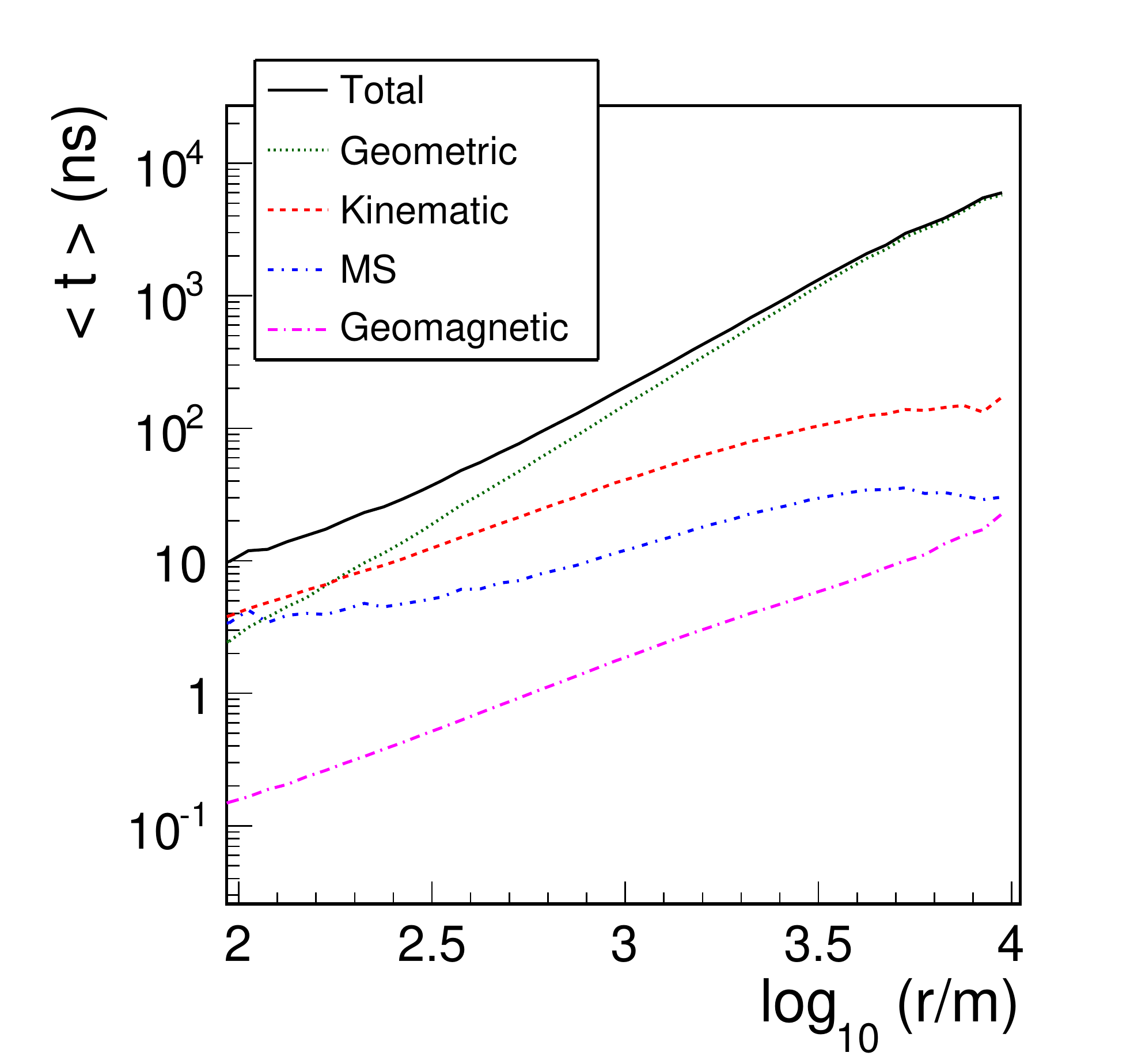}&
\includegraphics[width=9cm]{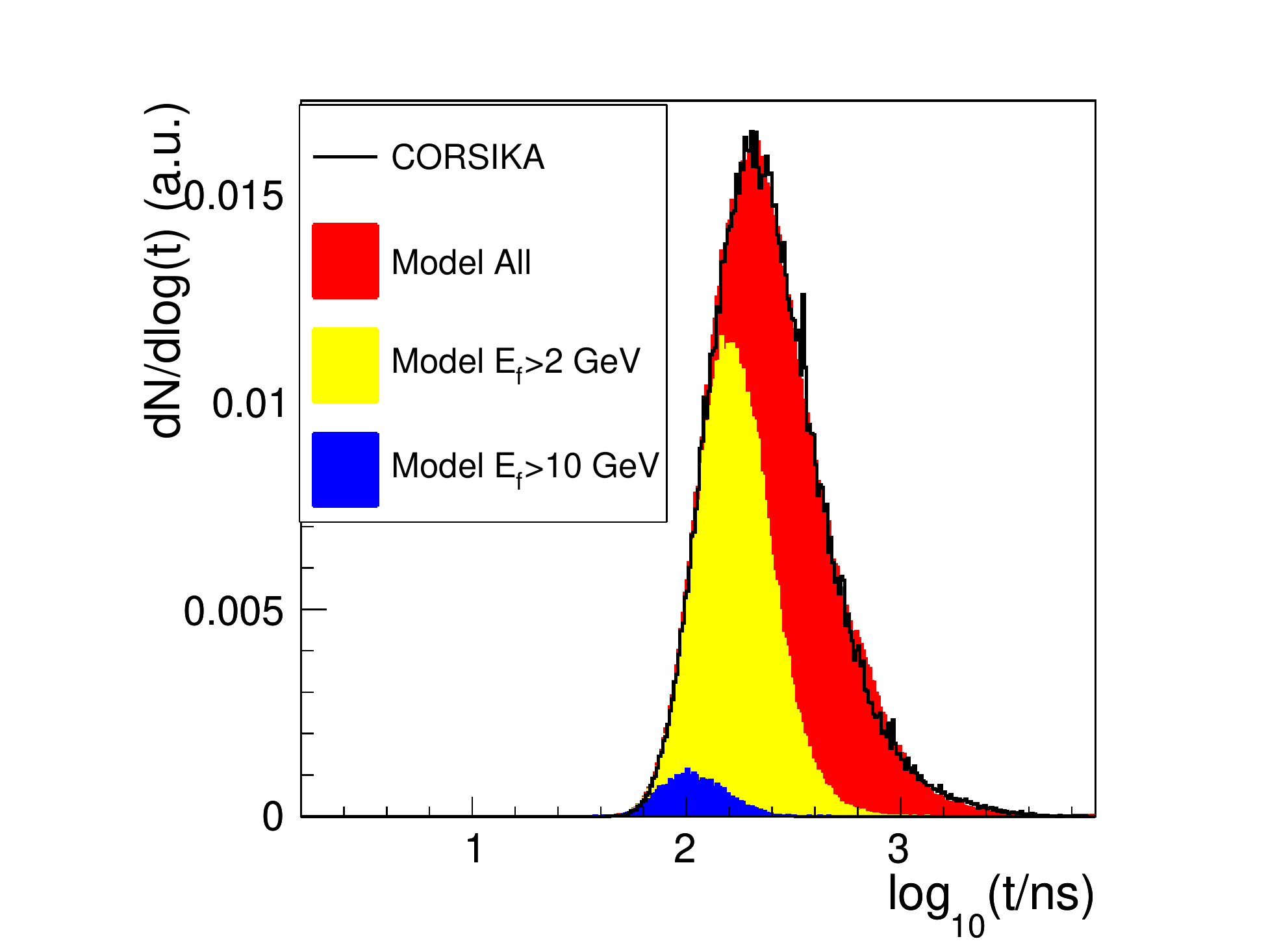}
\end{array}$
\end{center}
\caption{Left panel: Different contributions to the total average  time delay for a $60$  degrees shower. Right panel: Comparison between the model and CORSIKA of the normalized time distributions for a 60 degree shower $r=1300$ m distances from the core. The color histograms show the contribution of different energies.} 
\label{f:time}
\end{figure*}

The total time delay is the sum of four  different contributions $t=t_g+t_\epsilon+t_B+t_{MS}$
where $t_g$ is the geometric delay, $t_\epsilon$ is the kinematic delay, $t_B$ is the contribution produced by the geomagnetic field, and finally $t_{MS}$ includes  the delay due to multiple scattering.
Fig. \ref{f:time}, left panel, displays the different contributions to the total delay for $60$ degrees zenith angle.
At large distances from the core, the geometric delay is the most important. At distances typically from a few hundred meters to 1 km, the kinematic delay has a large impact. As we increase the zenith angle, the geometric delay looses importance relatively to the other contributions. At $500$ m from the core, the geometric delay represents $\simeq$60\% of the total.
Fig. \ref{f:time}, right panel, displays the overall time distributions at 1300 m from the shower core for a 60 deg shower. Filled histograms show the contributions of different muon energies at ground. High energy muons arrive earlier at ground. This is so because they are produced higher up in the atmosphere, and therefore have less geometric delay, but also because they have less kinematic delay.

The muon arrival time distributions can be used to extract relevant information. Far from the core, the time distributions are to a very good extent a one to one map of the {\it apparent} MPD-distributions. They can be determined by converting each muon time into a production distance, being the kinematic time a second order correction. Since the energy of each muon is typically not known, it is approximated by the mean value, taken from the energy spectrum at each observation point as it was explained in \cite{Cazon:2003ar,Cazon:2004zx,PhDCazon}. The energy would also determine the parameters of the multiple scattering delay distribution, although its concrete value follows a random distribution. The geomagnetic delay can take only two possible values depending on the charge of the muon. In general this technique will require a stringent $r$ cut for those regions where the geometric delay is a large fraction of the total delay, in order to avoid distortions of the reconstructed $dN/dX|_{(r,\zeta)}$. A more promising method consists in fitting the time distributions at once leaving a set of shape parameters on $h(X)$ free.  
Close to the core, the geometric delay is not dominant and the arrival time is mostly determined by the energy of each muon. This opens to possibility to measure, or at least constrain, the shape of the muon energy spectrum. A global fit would also allow to extract parameters from the $p_t$ distributions.

\subsection{Muon lateral distribution at ground}
The number of muons per surface area unit is $\rho(r,\zeta)=\frac{d^2N}{r dr d\zeta}$. As it was shown in \cite{Cazon:2012ti}, low energy muons have a major impact on the fine details of the muon lateral distribution at ground.

\begin{figure}[!t]
  \begin{center}
    \includegraphics[width=7cm]{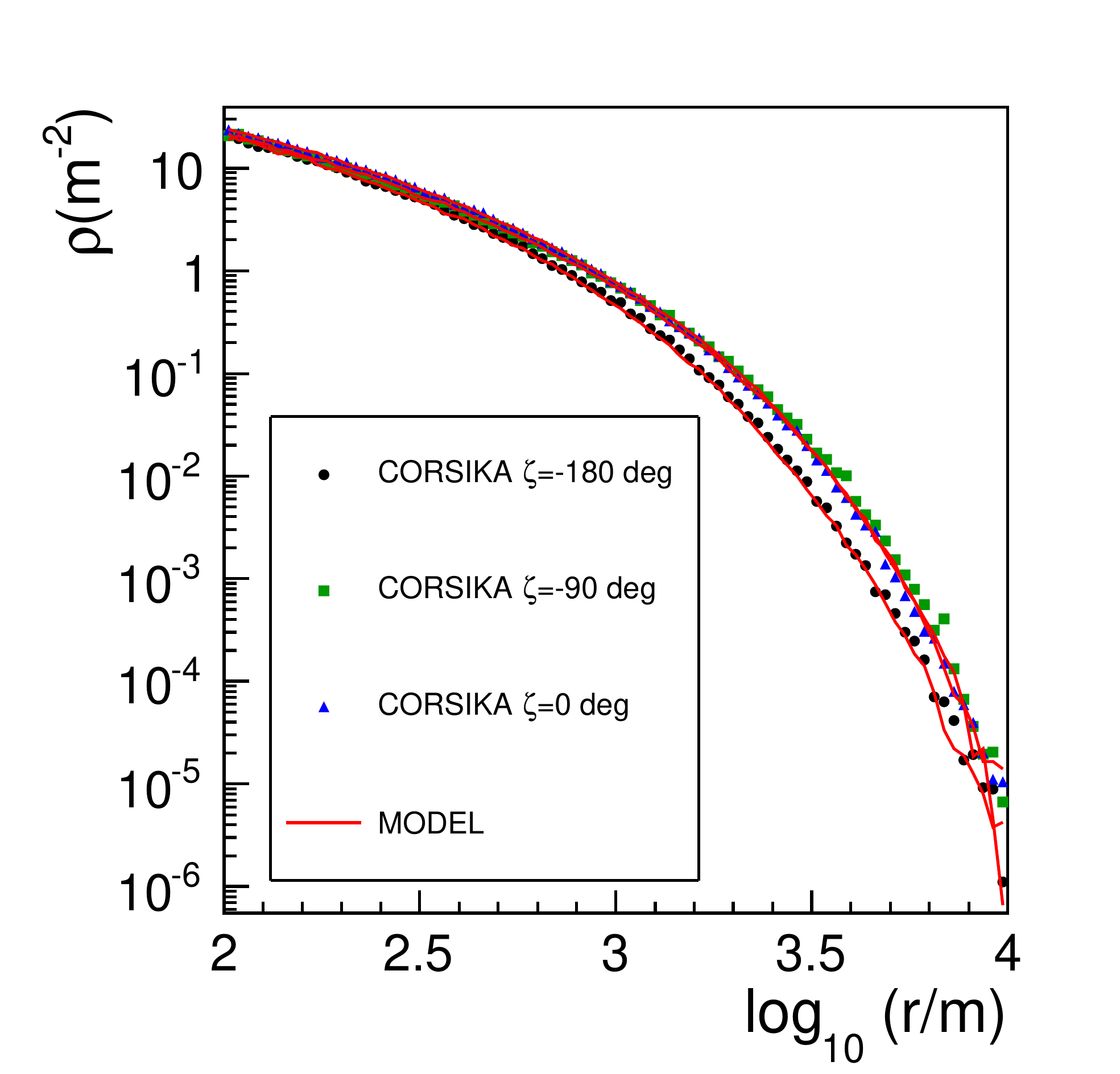}
\caption[]{Muon lateral distribution at ground for 3 different polar angles $\zeta$ for a proton shower at $10^{19}$ eV and 70 deg zenith angle.}
    \label{f:LDF}
 \end{center}
\end{figure}

 In vertical showers the number of muons per surface area does not depend much on $\zeta$. As we increase the zenith angle, asymmetries appear because of the different propagation effects, mainly decay and geometry.
The effects of the magnetic field become important above $60$ degrees, and they completely dominate the distributions at very inclined showers, typically between $80$ and $90$ degrees \cite{Ave:2000xs}. Fig. \ref{f:LDF} displays the muon density as a function of $r$ for 3 different polar angles $\zeta$ on a 70 deg shower.

\vspace{0.5cm}

The shape of the ground distributions is fully determined by the distributions at production, $h(X)$ and $f_X(E_i,p_t)$.  A change in the overall muon content of the shower, ${\cal N}_0$, produces a change in the muon density at ground, and therefore in the normalization of all distributions. The other main source of fluctuations comes from the depth of the first interaction, which directly affects $h(X)$ by changing its maximum, $\Xmax$. The position of $\Xmax$  directly influences all distributions at ground since it changes the total distance traveled by muons to ground.

\subsection{Average energy and transverse momentum distributions}

One of the main applications of the present model is to be used in a global fit to extract information on the total number of muons in the shower ${\cal N}_0$, and the {\it total/true} production depth distribution, $h(X)$, and its maximum, $\Xmax$. In order to do so,  a $f_X(E_i,p_t)$ distribution must be assumed.

The energy and transverse momentum distributions display more universal features when they are expressed in terms of $X'=X-\Xmax$, once the effects of the fluctuations induced by the first interaction point are removed.   The average energy and transverse momentum distributions do not change when changing the energy of the primaries, whereas they show mild differences between proton and iron primaries, and between hadronic interaction models.

If we substitute $f_X(E_i,cp_t)$ of a given shower by an average over showers of the same hadronic interaction model, primary, and zenith angle, $\left< f_{X'}(E_i,cp_t) \right>$, leaving only $h(X)$ from the original shower, the ground density displays differences of about $\sim$ 2\% at 1000 m compared to the prediction if we used $f_X(E_i,cp_t)$, whereas the rest of the ground distributions remained unchanged.  It is thus possible to use an universal energy and traverse momentum distribution that depends only on $X'$, where the position of $\Xmax$ is naturally accounted for through $X=X'+\Xmax$.

 The systematics of any concrete application, including a global fit, are to be studied and accounted for in each particular method and/or experimental setup. The effects of the choice of hadronic interaction model on $\left< f_{X'}(E_i,cp_t)\right>$ might introduce some systematics that should be also accounted for. On the contrary, those differences might be used to constrain $f_{X'}$ itself when compared to data which is very promising.  One could also think of a method to experimentally constrain the energy and transverse momentum spectrum based on simultaneous observations of the ground distributions in different conditions. For instance, the ground muon distributions of inclined showers contain valuable information about the energy spectrum due to the spectrographic effect of the geomagnetic field.

\section{Experimental efforts}
\label{sec-5}
In this section I will illustrate some of the experimental efforts to reconstruct the muon distributions with a few selected examples.

KASCADE has recently published \cite{Apel:2011zz} the {\it apparent} MPD-distributions for showers between E$\sim [10^{15},10^{17.7}]$ and zenith angle [0,18] deg at distances to the core [40,80] m. KASCADE uses a combination of different detectors which can separate the components of the shower, being possible to individually tag single muons. It also has a muon telescope, able to track the trajectory of the muon back to the shower axis and thus determine the production height. The back-tracking technique can be used in combination with the time-to-X technique in the Time-Track Complementarity method \cite{Ambrosio:1998gc}, which is able to separate high energy muons from low energy muons, opening new observables. As a drawback, these heavily instrumented observatories are hardly scalable to the high energy end of the spectrum, at energies around $10^{19}$ eV.

The Pierre Auger observatory has recently published the maximum of the  {\it apparent}-MPD using the time-to-X technique. Although the water Cherenkov tanks were not specifically designed to distinguish muons from electrons and photons, a fiducial cut can remove those stations close to the core and keep the muon richness sufficiently high to reconstruct the MPD-distributions.

KASCADE-Grande and Auger have also published in \cite{Souza:2011} and \cite{Allen:2011} the number of muons as a function of the energy, in two different energy ranges. Auger sees an excess in the number of muons when compared to simulations. It is still unclear whether the number of muons measured by the two experiments match due to the gap region around $10^{18}$ eV.

The shower-to-shower distribution of the number of muons contains also valuable information that it is not yet fully exploited. The RMS of the number of muons adds valuable information to help break the degeneracy between hadronic models and composition. To achieve this goal, it is important to enhance the muon capabilities at the highest energies and gain precision in the muon reconstruction. 

MARTA (Muon Auger RPC Tank Array) is one of the efforts in this direction. It envisages to add highly robust and autonomous RPC detectors \cite{Assis:2011} to the Auger Cherenkov tanks to enhance the muon capabilities.

\section{Conclusions}
\label{sec-6}
Particle reactions beyond the energy achieved by man made accelerators are continuously happening in the hadronic backbone of EAS.  Muons are true smoking guns of the hadronic cascade. In fact, we have shown that the relation between the distributions directly inherited from hadrons and the distributions observed at ground are well understood and that ground distributions can be used to reconstruct and constrain the distributions at production. We must study muons to enhance the sensitivity to the hadronic phenomena in the cascade. This is the path to solve the problem of composition of the UHECR and might uncover new particle physics phenomena at high energies. The future cosmic ray experiments above LHC energies need precise muon dedicated detectors added to the EAS arrays.

\section{Acknowledgments}
I would like to thank R.~\Conceicao, C.~Espirito-Santo, M.~Pimenta and M.~Oliveira for careful reading of this manuscript and their constructive comments. This work is partially funded by Funda\c{c}\~{a}o para a Ci\^{e}ncia e Tecnologia (CERN/FP11633/2010), and fundings of MCTES through POPH-QREN Tipologia 4.2, Portugal, and European Social Fund.

\end{document}